\documentstyle[epsf,float,seceq,twocolumn]{jpsj}

\relax

\author{Akihisa {\sc Koga}}
\title{
Ground-State Phase Diagram for the Three-Dimensional Orthogonal-Dimer System
}
\inst{Department of Applied Physics, Osaka University, 
Suita, Osaka 565-0871}
\recdate{\today}
\abst{
We investigate quantum phase transitions in the three-dimensional 
orthogonal-dimer system of the compound $\rm SrCu_2(BO_3)_2$ by means of 
the series expansion technique.
We then discuss the phase diagram where
the dimer phase, the Haldane phase, 
the frustration-induced disordered phase and 
the magnetically ordered phase compete with each other.
It is found that the compound $\rm SrCu_2(BO_3)_2$ is located 
in the dimer phase close to the phase boundary.
}
\kword{$\rm SrCu_2(BO_3)_2$, orthogonal-dimer, 
Shastry-Sutherland model, series expansion}
\def\runtitle{Effects of inter-layer coupling for the orthogonal-dimer system}
\def\runauthor{Akihisa {\sc Koga}}
\begin{document}
\sloppy
\maketitle
%%%%%%%%%%%%%%%%%%%%%%%%%%%%%%%%%%%%
%\section{Introduction}
%%%%%%%%%%%%%%%%%%%%%%%%%%%%%%%%%%%%
Two-dimensional (2D) quantum spin systems with a spin gap have been 
the subject of considerable interest.
In particular, the effect of the frustration in these systems 
plays an important role in stabilizing the disordered ground state.
The 2D spin-gap compound $\rm SrCu_2(BO_3)_2$ found recently\cite{Kageyama} 
has strong frustration due to its orthogonal-dimer structure,
which has provided a variety of interesting topics.
\cite{Onizuka,Nojiri,Room,Lemmens,Zherlitsyn,Kage2}
The system can be described by the 2D Heisenberg model on a square lattice with
some diagonal bonds, 
which is referred to as the Shastry-Sutherland model, \cite{Shastry}
as reported by Miyahara and Ueda.\cite{Miyahara}
In this 2D model, most of the fundamental properties including the ground state
as well as the excited states have been clarified by various theoretical 
approaches.\cite{Shastry,Miyahara,Ueda,3DMiyahara,Koga,Knetter,
MagFu,MagMi,MagMo,TwotriMo,Muller,TwotriFu,Mila,Weihong}
On the other hand, some recent reports claim that 
the compound $\rm SrCu_2(BO_3)_2$ has a large interlayer coupling,
which may not be negligible in explaining the experimental findings. 
\cite{3DMiyahara,Knetter}
Therefore it is necessary to study the phase diagram for 
the three-dimensional (3D) orthogonal-dimer system.

\begin{figure}[htb]
\vspace{0.1cm}
\epsfxsize=6.5cm
\centerline{\epsfbox{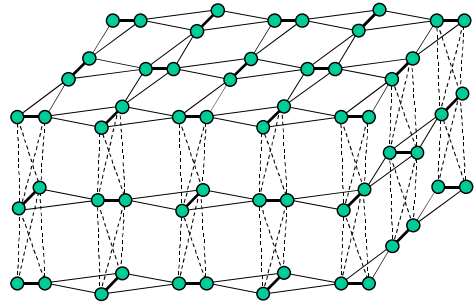}} 
\caption{The 3D orthogonal-dimer system for the compound $\rm SrCu_2(BO_3)_2$. 
The bold, solid and dashed lines indicate the antiferromagnetic exchange 
couplings $J$, $J'$ and $J''$, respectively.}
\label{fig:model}%-----------------------------------------
\end{figure}

\begin{figure}[htb]
\vspace{0.1cm}
\epsfxsize=6cm
\centerline{\epsfbox{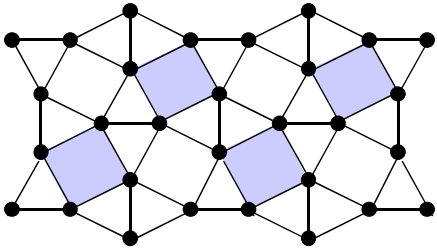}} 
\caption{The initial configuration of the plaquettes used
to treat the frustration-induced disordered phase by means of 
the series expansion technique.}
\label{fig:start1}%-----------------------------------------
\end{figure}

\begin{figure}[htb]
\vspace{0.1cm}
\epsfxsize=6cm
\centerline{\epsfbox{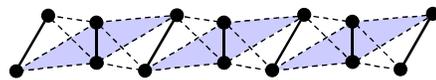}} 
\caption{The two-leg ladder with diagonal bonds,
which may be considered to be a chain of tetrahedra with shared edges.
\protect{\cite{Gelfand}}}
\label{fig:start2}%-----------------------------------------
\end{figure}

%%%%%%%%%%%%%%%%%%%%%%%%%%%%%%%%%%%
%\section{Model}\label{model}
%%%%%%%%%%%%%%%%%%%%%%%%%%%%%%%%%%%
Motivated by the above findings, 
we investigate here the quantum phase transitions 
for the 3D orthogonal-dimer system
which was introduced by Ueda and Miyahara. \cite{Ueda}
The Hamiltonian is described by 
\begin{eqnarray}
H=J\sum_{(i,j)}{\mib S}_i\cdot{\mib S}_j+
J'\sum_{(i,j)}{\mib S}_i\cdot{\mib S}_j+
J''\sum_{(i,j)}{\mib S}_i\cdot{\mib S}_j,\label{eq:H}
\end{eqnarray}
where ${\mib S}_i$ is the $S=1/2$ spin operator at the $i$-th site,
and $J$, $J'$ and $J''$ are antiferromagnetic couplings.
The structure of this model is shown schematically in Fig. \ref{fig:model}.
%%%%%%%%%%%%%%%%%%%%%%%%%%%%%%%%%%%%%
%%%%%%%%%%%%%%%%%%%%%%%%%%%%%%%%%%%%
In some limiting cases, the 3D model is reduced to
interesting orthogonal-dimer systems 
which have been studied intensively thus far.
When $J''=0$, the system is the well-known 
Shastry-Sutherland model\cite{Shastry} (see Fig. \ref{fig:start1}),
for which many groups have theoretically discussed various properties such as 
plateaus in the magnetization curve, 
\cite{Miyahara,Muller,MagFu,MagMi,MagMo,TwotriMo}
two-magnon excitations, \cite{TwotriFu,Knetter,TwotriMo}
and quantum phase transitions. 
\cite{Mila,Miyahara,Muller,Weihong,Koga,Knetter} 
In particular, the nontrivial disordered state 
adiabatically connected to isolated plaquettes\cite{Koga}
may be the most probable singlet state, which competes with the dimer state
as well as the antiferromagnetically ordered state.
In this paper, this phase is referred to as a "plaquette phase", 
for simplicity.
On the other hand, in the case $J'=0$,
the system is reduced to the two-leg ladders with the diagonal bonds
(see Fig. \ref{fig:start2}), 
which may be regarded as chains of the tetrahedra with shared edges, 
as pointed out by Gelfand. \cite{Gelfand}
In the ladder system, the dimer phase undergoes a first-order quantum phase
transition to another disordered phase, which is exactly formed by 
the triplet $(S=1)$ on each rung.
Note that the singlet on each rung is completely decoupled from 
the triplet for any value of $J''$, due to the orthogonal-dimer structure.
\cite{Gelfand,Kitatani,White,Lin,Honecker}
Therefore, this disordered phase is equivalent to the Haldane phase realized 
in the $S=1$ quantum Heisenberg chain. 
\cite{Haldane,Hida,Strong,Watanabe,Nishiyama,Narushima,Totsuka}
Concerning the first-order transition, 
it is known that the cusp and jump singularities appear 
in the excitations\cite{Gelfand,Lin} and 
the magnetization curve.\cite{Honecker}

In this study, we discuss the phase diagram
for the 3D system by means of the series expansion method.
\cite{series}
We observe how the frustration affects the quantum phase transitions.
It is also clarified that the spin-gap compound 
$\rm SrCu_2(BO_3)_2$ is located in the vicinity of the phase boundary 
which separates the dimer and the magnetically ordered phases.

%%%%%%%%%%%%%%%%%%%%%%%%%%%%%%
%\section{Quantum phase transitions}
%%%%%%%%%%%%%%%%%%%%%%%%%%%%%%
In the series expansion approach,\cite{first,series} 
we start with isolated clusters of a proper spin-singlet configuration.
For this purpose, we divide the original Hamiltonian into 
two parts as $H=H_0+\lambda H_1$,
where $H_0 (H_1)$ represents the unperturbed (perturbed) Hamiltonian and
$\lambda$ is an auxiliary parameter.
Details of the initial configurations for various phases are given 
in the following.
By turning on the couplings among the clusters perturbatively $(H_1)$,
we carry out the series expansion in $\lambda$ 
for the ground state energy and the spin gap.
Then the critical point for the first- (second-)order transition 
is determined by applying the first-order inhomogeneous differential method 
(Pad\'e approximants) to the obtained series. \cite{Pade}
In the following, starting with the Shastry-Sutherland model and 
the two-leg ladder model with diagonal bonds,
we deal with quantum phase transitions for the 3D orthogonal-dimer model.

%\subsection{Quasi-2D Shastry-Sutherland model}
%Let us first discuss how the inter-layer coupling 
%on the 2D Shastry-Sutherland model $(J''=0)$ 
%affects the quantum phase transitions.

\begin{figure}[htb]
\vspace{0.1cm}
\epsfxsize=8cm
\centerline{\epsfbox{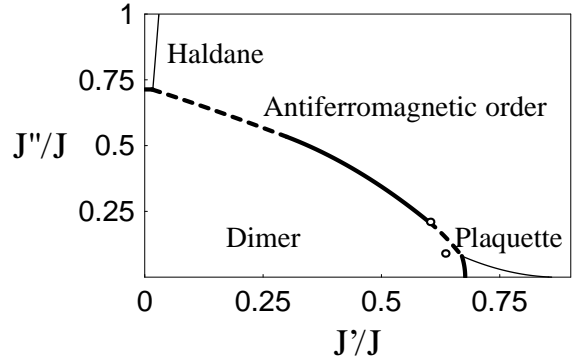}} 
\caption{The phase diagram for the 3D orthogonal-dimer system.
The bold and solid lines represent the phase boundaries where the first-
and the second-order quantum phase transitions occur, respectively.
The open circles indicate the locations of the orthogonal-dimer 
compound $\rm SrCu_2(BO_3)_2$ obtained by 
Miyahara and Ueda,\protect{\cite{3DMiyahara}}
and Knetter et al.\protect{\cite{Knetter}}
}
\label{fig:phase}%-----------------------------------------
\end{figure}

Note that the direct product of the dimer-singlet states 
formed by the couplings $J$ is always an eigenstate of the Hamiltonian 
(\ref{eq:H}) with the energy $E=-3/8 NJ$, even in the 3D model, \cite{Ueda}
where $N$ is the total number of spins.
We then discuss how the interlayer coupling 
on the 2D Shastry-Sutherland model $(J''=0)$ 
affects the quantum phase transitions.
We recall here the scenario in our previous paper\cite{Koga}
where the first- (second-)order quantum phase transition in 2D
was discussed by calculating the ground state energy (the spin gap) 
for the plaquette phase.
Following this concept, we compute the quantities for the plaquette phase 
by means of the series expansion technique.
We choose the arrangement of plaquettes 
shown in Fig. \ref{fig:start1} as a starting configuration.
%%%%%%%%%%%%%%%%%%%%%%%%%%%%%%%%%%%%%
%%%%%%%%%%%%%%%%%%%%%%%%%%%%%%%%%%%%
In this figure, the shaded squares represent the plaquettes formed 
by the exchange coupling $J'$.
The other bold and solid lines indicate 
the exchange couplings between the plaquettes, 
$\lambda J$ and $\lambda J'$, respectively.
We also set the interlayer coupling as $\lambda J''$, 
which is not shown in this figure, for simplicity.
To discuss the first-order quantum phase transition to the dimer phase,
we perform the plaquette expansion for the ground state energy 
up to the fifth order in $\lambda$ for several values of $J'$ and $J''$.
We estimate the ground state energy in the case $\lambda=1$, where 
the model is reduced to the original orthogonal-dimer model
by means of the first-order inhomogeneous differential method.\cite{Pade}
Comparing the energies for the dimer and the plaquette phases, 
we determine the phase boundary shown as the bold line in Fig. \ref{fig:phase},
which separates two disordered phases. 
%%%%%%%%%%%%%%%%%%%%%%%%%%%%%%%%%%%%%
%%%%%%%%%%%%%%%%%%%%%%%%%%%%%%%%%%%%
When $J''=0$ (Shastry-Sutherland model), the plaquette phase undergoes
the first-order quantum phase transition at $(J'/J)_c=0.68$.
Since this critical value obtained by the fifth-order expansion is consistent 
with the result $(J'/J)_c=0.677$ obtained 
by the seventh-order expansion,\cite{Koga}
it is expected that this lower-order expansion with the first-order 
inhomogeneous differential method may give the phase boundary 
quantitatively well even in the 3D model.
By introducing the interlayer coupling $J''$, 
the energy for the plaquette phase seems to decrease quadratically,
while the energy for the dimer phase is not 
changed due to its orthogonal-dimer structure.
This implies that the phase boundary between the two spin-gap phases
is little affected as long as the interlayer couplings are small.

In order to completely discuss the interlayer effects,
the antiferromagnetically ordered phase stabilized by 
the interlayer coupling must be properly taken into account.
Performing the plaquette expansion for the spin gap up 
to the fourth order in $\lambda$, 
we determine the second-order quantum 
phase transition point by means of the biased Pad\'e approximants 
with the critical exponent $\nu\sim0.6$. 
The introduction of the interlayer couplings induces
the antiferromagnetic correlation, 
while it suppresses the frustration which stabilizes the plaquette phase.
%It is found that the region of the plaquette phase is very small 
%in the phase diagram for the 3D model.
The multi-critical point can be estimated as
$(J'/J, J''/J)\sim(0.67,0.08)$, and
it is seen that the plaquette phase induced by the frustration is not 
stable against the interlayer coupling.

%\subsection{Quasi-1D two-leg ladder system}
We next consider the present system as the coupled two-leg ladders
\cite{Gelfand} $(J'\sim0)$, as shown in Fig. \ref{fig:start2}.
%%%%%%%%%%%%%%%%%%%%%%%%%%%%%%%%%%%%%
%%%%%%%%%%%%%%%%%%%%%%%%%%%%%%%%%%%%
We discuss how the dimer and the Haldane phases compete 
with the magnetically ordered phase induced by the inter-ladder coupling $J'$.
Since the Haldane phase is formed by the triplet state ($S=1$) 
on each rung, the dimer expansion starting from the singlet state ($S=0$)
is not appropriate to discuss the critical phenomena.
Nevertheless, following the concept of the valence bond solid, \cite{VBS}
we can choose a suitable initial configuration for the Haldane phase 
in the framework of the series expansion. \cite{KogaHaldane}
Namely, we introduce the unperturbed Hamiltonian $H_0$,
by choosing the initial configuration of isolated plaquettes 
indicated by the shaded rectangles in Fig. \ref{fig:start2}.
Considering the other couplings as perturbations,
we then discuss the stability of the Haldane phase.
To study the first- (second-)order transition, 
we perform the plaquette expansion for 
the ground state energy (the spin gap) up to the sixth (fifth) order 
in $\lambda$.
Using the approximation discussed above,
we then determine the phase boundaries shown in Fig. \ref{fig:phase}.
When $J'=0$, the first-order quantum phase transition 
from the Haldane phase to 
the dimer phase occurs at the critical value $(J''/J)_c \sim 0.71$, 
which is in good agreement with the accurate value 
$(J''/J)_c = 0.7135(3)$ obtained by Gelfand.\cite{Gelfand}
Introducing the interladder coupling $J'$ in the Haldane phase 
($J''/J>0.71$), the antiferromagnetic correlation is enhanced and 
the spin gap decreases rapidly.
Finally, the second-order quantum phase transition occurs 
to the magnetically ordered phase. 
We estimate the multicritical point $(J'/J, J''/J)\sim(0.015,0.7)$
where the dimer, the Haldane and the magnetically ordered phases 
compete with each other.

Finally, to discuss the competition between the dimer and 
the antiferromagnetically ordered phases,
we use the Ising expansion technique.
To this end, we divide the original Heisenberg Hamiltonian into 
two parts as $H=\sum J_{ij} S^z_i S^z_j+ \lambda \sum J_{ij} (S^x_i S^x_j+
S^y_i S^y_j)$, where $J_{ij}$ is the exchange coupling for each bond.
The ground state energy $E_{\rm AF}$ for the antiferromagnetically 
ordered phase is given by 
\begin{eqnarray}
E_{\rm AF}&=&\frac{1}{2}NJS^2-2NJ'S^2-2NJ''S^2+\epsilon(\lambda),\label{eq:SW}
\end{eqnarray}
where $\epsilon(\lambda)$ is the quantum correction due to the $XY$ components.
Performing the Ising expansion for $\epsilon(\lambda)$ 
up to the fourth order in $\lambda$ and 
applying the Pad\'e approximants\cite{Pade} to the obtained series,
we then deduce the energy $E_{\rm AF}$.
By comparing it with the energy for the dimer phase,
we can determine the phase boundary, 
which is shown as the bold line in Fig. \ref{fig:phase}.
Near the multicritical points, 
the energy $E_{\rm AF}$ has not been deduced 
accurately, and thereby the phase boundary shown as the broken line is 
to guide the eyes.
When $J'=J''=0$, the 3D orthogonal-dimer system belongs to the dimer phase
with the exact dimer-singlet state.
Note that in the dimer phase the interlayer coupling $J''$ has little effect
on the ground state properties due to its orthogonal-dimer structure,
which is quite different from the case of the coupling $J'$.\cite{3DMiyahara}
Therefore the introduction of the interdimer couplings $J'$ and $J''$
enhances the 2D spin-spin correlation, and finally 
induces the first-order quantum phase transition
to the antiferromagnetically ordered phase with the 3D structure, 
as seen in Fig. \ref{fig:phase}.

Recently it was clarified that the compound $\rm SrCu_2(BO_3)_2$ 
has a large interlayer coupling.
In the reports by Miyahara and Ueda\cite{3DMiyahara} 
and Knetter et al.\cite{Knetter}, the ratios of the interdimer and 
interlayer exchange couplings are estimated as $(J'/J, J''/J)=
(0.635, 0.09)$ and $(0.603, 0.21)$,
which are indicated by open circles in Fig. \ref{fig:phase}.
It is found that the compound $\rm SrCu_2(BO_3)_2$ may be located 
in the dimer phase close to the phase boundary.

Some comments are in order for the possibility of 
another phase induced by the frustration in the ground state.
Recently Knetter et al. have found 
that the instability of the two-magnon excitation 
in the Shastry-Sutherland model $(J''=0)$ exists 
in the dimer phase.\cite{Knetter}
It is also claimed that the first-order quantum phase transition occurs 
to the classical helical ordered phase at the critical point 
$(J'/J)_c=0.630(5)$.
This result appears to contrast with 
the other results\cite{Miyahara,Koga,Weihong,Muller} 
except for those obtained by the Schwinger boson approach.\cite{Mila}
Although their conclusion of the first-order transition in terms of the
analysis of elementary excitations is speculative, it is an interesting
open problem to clarify what this instability really implies,
which is now under consideration.

%%%%%%%%%%%%%%%%%%%%%%%%%%%%%%%%%%%%
%\section{Summary}
%%%%%%%%%%%%%%%%%%%%%%%%%%%%%%%%%%%%
In summary, we have investigated the quantum phase transitions 
in the 3D orthogonal-dimer system for the compound $\rm SrCu_2(BO_3)_2$
by means of the plaquette expansion technique.
We have thus determined a phase diagram with a rich structure,
in which it is clarified that the frustration-induced disordered phase
is unstable against the interlayer couplings.
Also it is found that the compound $\rm SrCu_2(BO_3)_2$ is located 
in the dimer phase close to the phase boundary.

%%%%%%%%%%%%%%%%%%%%%%%%%%%%%%%%%%%%
%\acknowledgements
%%%%%%%%%%%%%%%%%%%%%%%%%%%%%%%%%%%%
We would like to thank N. Kawakami and  K. Okunishi 
for useful discussions. 
The work is partly supported by a 
Grant-in-Aid from the Ministry of Education, Science, Sports, 
and Culture. 
The author is supported by the Japan Society 
for the Promotion of Science. 
Some of the numerical computations in this work were carried out 
at the Yukawa Institute Computer Facility.

\end{document}